\documentclass[prb,aps,10pt,twocolumn,showpacs,floats,amsmath,amssymb,longbibliography]{revtex4-1}

\usepackage{hyperref}
\usepackage{graphicx}
\usepackage{dcolumn}
\usepackage{bm}
\usepackage{amssymb}
\usepackage{subfigure}
\usepackage{color}

\def \be{\begin{equation}}
\def \ee{\end{equation}}

\begin{document}

\title{Using Activated Transport in Parallel Nanowires \\ for Energy Harvesting and Hot Spot Cooling}

\author{Riccardo Bosisio}
\altaffiliation{Present address: NEST, Istituto Nanoscienze, Piazza San Silvestro 12, 56127 Pisa, Italy}
\author{Cosimo Gorini}
\altaffiliation{Present address: Institut f\"{u}r Theoretische Physik, Universit\"{a}t Regensburg, 93040 Regensburg, Germany}
\author{Genevi\`eve Fleury}
\author{Jean-Louis Pichard}
\affiliation{Service de Physique de l'\'Etat Condens\'e, DSM/IRAMIS/SPEC, CNRS UMR 3680, CEA Saclay, 91191 Gif-sur-Yvette Cedex, France}

\begin{abstract}
We study arrays of parallel doped semiconductor nanowires in a temperature range where the electrons propagate through the nanowires by phonon assisted hops 
between localized states. By solving the Random Resistor Network problem, we compute the thermopower $S$, the electrical conductance $G$, and the 
electronic thermal conductance $K^e$ of the device. We investigate how those quantities depend on the position -- which can be tuned with a back 
gate -- of the nanowire impurity band with respect to the equilibrium electrochemical potential. We show that large power factors  can be reached 
near the band edges, when $S$ self-averages to large values while $G$ is small but scales with the number of wires. Calculating the amount of heat 
exchanged locally between the electrons inside the nanowires and the phonons of the environment, we show that phonons are mainly absorbed near one 
electrode and emitted near the other when a charge current is driven through the nanowires near their band edges. This phenomenon could be exploited 
for a field control of the heat exchange between the phonons and the electrons at submicron scales in electronic circuits. It could be also used for 
cooling hot spots.
\end{abstract}


\maketitle
\section{Introduction}
\begin{figure}
    \includegraphics[clip,keepaspectratio,width=\columnwidth]{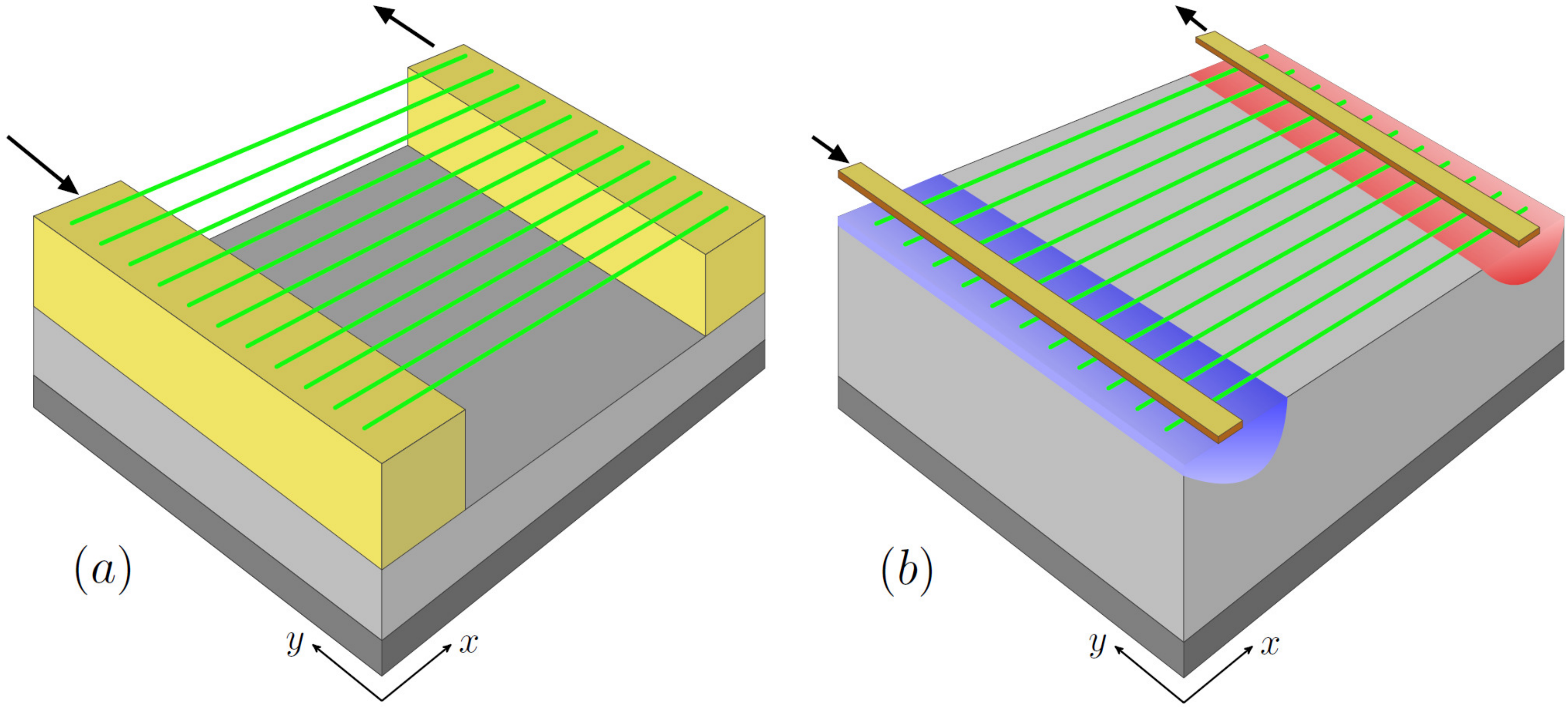}
    \caption{(Color online) Array of suspended (a) and deposited (b) parallel NWs in the FET configuration. The NWs are drawn in green, 
the two metallic electrodes in yellow, the substrate in grey and the back gate in dark grey. The blue [red] strip in (b) indicates the 
substrate region that is cooled down [heated up] in the phonon-assisted activated regime, when a charge current flows from the left to the 
right electrode and the gate voltage is tuned so as to probe the lower edge of the NWs impurity band.}
    \label{fig:sys}
\end{figure}

A good thermoelectric machine must be efficient at converting heat into electricity and also must provide a substantial electric output 
power for practical applications. In the linear response regime, this requires optimizing simultaneously the figure of merit 
$ZT=S^2GT/(K^e+K^{ph})$ and the power factor $\mathcal{Q}=S^2G$, $T$ being the operating temperature, $S$ the device thermopower, 
$G$ its electrical conductance, and $K^e$ and $K^{ph}$ its electronic and phononic thermal conductances.
In the quest for high performance thermoelectrics, semiconductor nanowires (NWs) are playing a front 
role,\cite{Hicks1993,Curtin2012,Blanc2013,Brovman2013,Stranz2013,Karg2013} apparently offering the best of three worlds.  
First, an enhanced $S$ due to strongly broken and gate-tunable particle-hole symmetry.\cite{Brovman2013,Roddaro2013,Moon2013,Tian2012}
Second, a suppressed $K^{ph}$ by virtue of reduced dimensionality.\cite{Curtin2012,Blanc2013} Finally, a high power output thanks to 
scalability, i.e. parallel stacking.\cite{Hochbaum2008,Curtin2012,Stranz2013,Atashbar2004,Yerushalmi2007,Wang2009,Zhang2010,Davila2011,Farrell2012,Pregl2013}\\
\indent
The perspective of developing competitive thermoelectric devices with the standard building blocks of the semiconductor industry has raised a great interest 
in the scientific community over the last decade. On a technological standpoint, much effort has been put into the synthesis of dense NWs arrays with controlled 
NW diameter, length, doping, and crystal orientation.\cite{Atashbar2004,Yerushalmi2007,Wang2009,Persson2009,Zhang2010,Farrell2012,Pregl2013}
Arrays made out of various semiconductor materials including e.g. Silicon, Silicon Germanium, Indium Arsenide, or Bismuth Telluride have thus been investigated.
Versatile measurement platforms have been developed to access the set of thermoelectric coefficients and the feasibility of NW-based thermoelectric modules
have been assessed.\cite{Abramson2004,Keyani2006,Hochbaum2008,Davila2011,Curtin2012,Stranz2013} 
On the theory side, numerous calculations of $S$, $G$, $K^e$ and $K^{ph}$ of various single NWs have been carried out in the ballistic regime of electronic 
transport~\cite{ODwyer2006,Markussen2009,Liang2010,Gumbs2010,Neophytou2010,Wang2014} or in the diffusive regime~\cite{Markussen2009} where a 
semi-classical Boltzmann approach can be used.
\cite{Lin2000,Humphrey2005,Bejenari2008,Vo2008,Shi2009,Bejenari2010,Neophytou2012,Ramaya2012,Neophytou2014,Bejenari2014,Curtin2014,Davoody2014} 
In two recent works,\cite{Bosisio20141,Bosisio20142} we took a different approach by considering the presence of electronic localized states randomly 
distributed along the NWs and making up an impurity band in the semiconductor band gap. Such states are known to play a leading role in thin nanowires, 
where localization effects are enhanced by low dimensionality and the system size rapidly exceeds the electron localization length. After a first study devoted 
to the low temperature coherent regime,\cite{Bosisio20141} we investigated the phonon-assisted hopping regime~\cite{Bosisio20142} taking place at higher 
temperatures and usually referred to as Mott activated regime. For a long time, thermoelectric transport in this regime has been somewhat overlooked in the 
theoretical literature (with the exception of a few older works on bulk semiconductors\,\cite{Zvyagin1973,Zvyagin1991,Movaghar1981,Wysokinski1985}). In fact, 
the problem of thermally-activated thermoelectric transport in NWs has been revisited only recently by Jiang \textit{et al.} in Refs.~\cite{Jiang2012,Jiang2013}. 
However the case of gated NWs where band edges are approached has not been considered though band-edge transport, where particle-hole asymmetry is maximal, 
is acknowledged to be the critical one for thermoelectric conversion~\cite{Mahan1989,Shakouri2011}. In our previous paper,\cite{Bosisio20142} we studied 
the behavior of the thermopower $S$ and of the electrical conductance $G$ of single disordered and gated NWs in the activated regime. We obtained near 
the band edges a substantial enhancement of the typical thermopower $S_0$ but also, unsurprisingly, a decrease of the typical conductance $G_0$ and large 
sample-to-sample fluctuations of both $G$ and $S$. This is unsatisfactory if a reliable and efficient thermoelectric device is to be realized.\\
\indent In the present paper, we  circumvent the latter shortcomings by considering a large set of NWs stacked in parallel in the field effect transistor (FET) 
device configuration. Besides assessing the opportunities offered by band edge activated transport for energy harvesting, we show that activated transport through such a device 
can be used for an electrostatic control of the heat exhange between the phonons and the electrons at sub-micron scales: Injecting the carriers through 
the NWs gives rise to a local cooling [heating] effect near the source [drain] electrode when the chemical potential of the device probes the lower NWs band edge (and conversely when it probes the upper edge). This opens promising perspectives for a local management of heat and for cooling hot-spots in microelectronics.\\
\indent Hereafter, we study arrays of doped semiconductor NWs, arranged in parallel and attached to two electrodes. The NWs can be either 
suspended or deposited onto an electrically and thermally insulating substrate. A metallic gate beneath the sample is used to 
vary the carrier density inside the NWs. This corresponds to a setup in the FET configuration, as sketched in Fig.~\ref{fig:sys}. If the thermopower 
or the thermal conductances are to be investigated, a heater (not shown in Fig.~\ref{fig:sys}) is added on one side of the sample to induce a temperature 
gradient between the electrodes. We focus on a temperature range where the activated regime 
proposed by Mott [Variable Range Hopping (VRH) regime] takes place, assuming \textit{(i)} that phonon-assisted transport occurs between localized 
states of the NWs impurity band only and \textit{(ii)} that the substrate, or the NWs themselves if they are suspended, act as a phonon bath to which 
NWs charge carriers are well coupled.  We thus consider intermediate temperatures, where the thermal energy $k_BT$ is high enough to allow inelastic 
hopping between Anderson localized states of different energies (typically a few Kelvin degrees), yet low enough to keep localization effects. 
Such VRH regime is observed up to room temperatures in three-dimensional amorphous semiconductors~\cite{Shklovskii1984} and very likely up to 
higher temperatures in the one-dimensional limit.  Following Refs.~\cite{Jiang2012,Jiang2013,Bosisio20142}, we solve numerically the Miller-Abrahams 
Random Resistor Network problem~\cite{Miller1960} for obtaining $S$, $G$, and $K^e$. This allows us to identify also the regions where heat exchanges 
between the electrons and the phonons dominantly take place in the activated regime, notably when the chemical potential probes the edges of the 
NWs impurity band.\\ 
\indent The model used throughout the paper is presented in Sec.~\ref{sec_method}, together with a summary of the method. We find in 
Sec.~\ref{sec_scaling} that once a large set of NWs is stacked in parallel, the strong $G$, $S$ and $K^e$ fluctuations are suppressed.
Denoting by $G_0$, $S_0$ and $K^e_0$ the {\it typical} values for a single NW, we observe more precisely that the thermopower of a large 
NW array self-averages ($S\to {S_0}$) while its electrical and electronic thermal conductances $G\to M {G_0}$, $K^e\to M {K^e_0}$ as the 
number $M$ of wires in parallel increases (see Fig.~\ref{fig:fluctuations}). Taking full advantage of the gate, we move close to the impurity band 
edges, where we recently obtained a drastic ${S_0}$ enhancement.\cite{Bosisio20142} We show in Sec.~\ref{sec_QZT} that in this regime a large 
$S_0$ partly compensates an exponentially small $G_0$, so that substantial values of the power factor ${\mathcal Q}\approx MS_0^2G_0$ can be reached 
upon stacking plenty of NWs in parallel [see Fig.~\ref{fig:QZT}(a)]. Remarkably, the electronic figure of merit $Z_eT=S^2GT/K^e$ is also found to 
reach promising values $Z_eT\approx 3$ when ${\mathcal Q}$ is maximal [see Fig.~\ref{fig:QZT}(b)]. Furthermore, we discuss how the phononic thermal 
conductance $K^{ph}$ will inevitably reduce the full figure of merit $ZT$ and argue that, even if record high $ZT$ is probably not to be sought in 
such setups, the latter have the great advantage of offering at once high output power and reasonable efficiency with standard nanotechnology building 
blocks. 
The most important result of this paper is given in Sec.~\ref{sec_hotspots}. 
We study how deposited NWs in the FET configuration can be used to manage heat in the substrate, generating hot/cold spots ``on demand''. The idea is 
simple to grasp and relies on the calculation of the local heat exchanges between the NWs electrons and the substrate phonons: When the gate voltage is 
adjusted such that the equilibrium electrochemical potential $\mu$ (defined in the electronic reservoirs) roughly coincides with one (say the lower) 
impurity band edge, basically all energy states in the NWs lie above $\mu$.  Therefore, if charge carriers injected into the system around $\mu$ are 
to gain the other end, they need to (on the average) absorb phonons at the entrance so as to jump to available states, and then to release phonons 
when tunneling out (again at $\mu$). This generates in the nearby substrate regions cold strips near the injecting electrode and hot strips near the 
drain electrode [see Figs.~\ref{fig:sys}(b) and~\ref{fig:hotspots}]. These strips get scrambled along the nanowires if $\mu$ does not probe the edges of the NWs impurity band. 
Such reliable and tunable cold spots may be exploited in devising thermal management tools for high-density circuitry, where ever increasing power densities 
have become a critical issue.\cite{Vassighi2006} Moreover, the creation/annihilation of the cold/hot strips can be controlled by the back gate voltage.
Note that the underlying mechanism governing the physics of VRH transport at the NWs band edges is somewhat reminiscent of the mechanism of "cooling by 
heating" put forward in Refs.~\cite{Pekola2007,Rutten2009,Levy2012,Marl2012,Cleuren2012}, which also exploits the presence of a third bosonic bath in 
addition to the two electronic reservoirs. In our case, bosons are phonons provided mainly by the substrate; in other setups, bosons are photons provided 
by laser illumination (or more simply by the sun for a photovoltaic cell). All those studies fall into the growing category of works dealing with 
boson-assisted electronic transport that have been shown to open promising perspectives for heat management.

\section{Model and method}
\label{sec_method}
Architecture and/or material specific predictions, though very important for practical engineering purposes, are however {\it not} our concern at present. 
On the contrary, our goal is to reach conclusions which are as general as possible, relying on a bare-bone but widely applicable Anderson model devised to 
capture the essentials of the physics we are interested in. We consider a set of $M$ NWs in parallel. Each NW is modeled as a chain of length $L$ described 
by a one-dimensional (1D) Anderson tight-binding Hamiltonian with on-site disorder~\cite{Bosisio20142}:
\be
\label{eq_modelAnderson1D}
\mathcal{H}=-t\sum_{i=1}^{N-1}\left(c_i^{\dagger}c_{i+1}+\text{h.c.}\right)+\sum_{i=1}^{N}(\epsilon_i+V_g) c_i^{\dagger}c_i\,.
\ee
Here $N$ is the number of sites in the chain ($L=Na$ with $a$ lattice spacing), $c^{\dagger}_i$ and $c_i$ are the electron creation and 
annihilation operators on site $i$ and $t$ is the hopping energy ({\it inter}-wire hopping is neglected). We assume that no site can be doubly 
occupied due to Coulomb repulsion, but otherwise neglect interactions.\cite{Ambegaokar1971} The site energies $\epsilon_i$ are uncorrelated 
random numbers uniformly distributed in the interval $[-W/2,W/2]$, while $V_g$ is a constant (tunable) potential due to the back gate. 
The electronic states are localized at certain positions $x_i$ with localization lengths $\xi_i$ and eigenenergies $E_i$. The $E_i$'s lie within 
the NW impurity band whose center can be shifted with the gate voltage $V_g$. For simplicity's sake, we generate randomly the positions 
$x_i$ along the chain (with a uniform distribution) and assume $\xi_i=\xi(E_i)$, where $\xi(E)$ characterizes the exponential decay of the 
\textit{typical} conductance $G_0\sim\exp(-2L/\xi)$ of the 1D Anderson model at zero temperature and energy $E$. Analytical expressions giving $\xi(E)$ 
in the weak disorder limit of the Anderson model are given in Ref.~\cite{Bosisio20141}.  

\indent The NWs are attached to two electronic reservoirs $L$ and $R$, and to a phonon bath, i.e. the system is in a three-terminal configuration. 
Particles and heat(energy) can be exchanged with the electrodes, but only heat(energy) with the phonon bath. At equilibrium the whole system is 
thermalized at a temperature $T$ and both $L$ and $R$ are at electrochemical potential $\mu$ (set to $\mu\equiv 0$, at the band center 
when $V_g=0$). A voltage and/or temperature bias between the electrodes drives an electron current through the NWs. Hereafter we consider the 
linear response regime, valid when small biases $\delta\mu\equiv\mu_L-\mu_R$ and $\delta T\equiv T_L-T_R$ are applied.

\indent We study the inelastic activated regime. Following ref.\cite{Jiang2013}, we assume that the charge carriers (say electrons of 
charge $e$) tunnel elastically from reservoir $\alpha=L,R$ into some localized states $i$ whose energies $E_i$ are located in a window of 
order $k_BT_\alpha$ around $\mu_\alpha$. They then proceed via phonon-assisted hops to the other end, finally tunneling out. The maximal 
carriers hop along the NWs is of the order of Mott length $L_M$ in space (or Mott energy $\Delta$ in energy)~\cite{Bosisio20142}. At the 
lowest temperatures considered in this work, $\xi(\mu)\ll L_M\ll L$ and transport is of Variable Range Hopping (VRH) type.  An increasing 
temperature shortens $L_M$ until $L_M\approx\xi(\mu)$, when the Nearest Neighbors Hopping (NNH) regime is reached. The crossover 
VRH$\rightarrow$NNH takes place roughly at Mott temperature $T_M$, whose dependence on $V_g$ can be found in Ref.~\cite{Bosisio20142}.

\indent The \textit{total} electron and heat currents flowing through the \textit{whole} array are calculated by solving the Random Resistor 
Network problem.\cite{Miller1960,Ambegaokar1971} The method is summarized in Appendix \ref{app_rrn}. It takes as input parameters the rate 
$\gamma_e$ quantifying the coupling between the NWs (localized) and the reservoirs (extended) states, and the rate $\gamma_{ep}$ measuring the 
coupling to the NWs and/or substrate phonons.  We point out that we go beyond the usual approximation~\cite{Miller1960,Ambegaokar1971,Jiang2013} 
neglecting the $\xi_i$'s variations from state to state [$\xi_i\approx\xi(\mu)$], the latter being inappropriate close to the band edges, where 
$\xi_i$ varies strongly with the energy. Following Ref.~\cite{Bosisio20142} the random resistor network is then solved for $\xi_i\neq\xi_j$. 
The particle and heat currents thus obtained are related to the small imposed biases $\delta\mu, \delta T$ via the Onsager matrix,\cite{Callen1985} 
which gives access to $G$, $K^e$ and $S$.
\section{Scaling of the thermoelectric coefficients with the number of nanowires}
\label{sec_scaling}
\begin{figure}
\includegraphics[keepaspectratio,width=\columnwidth]{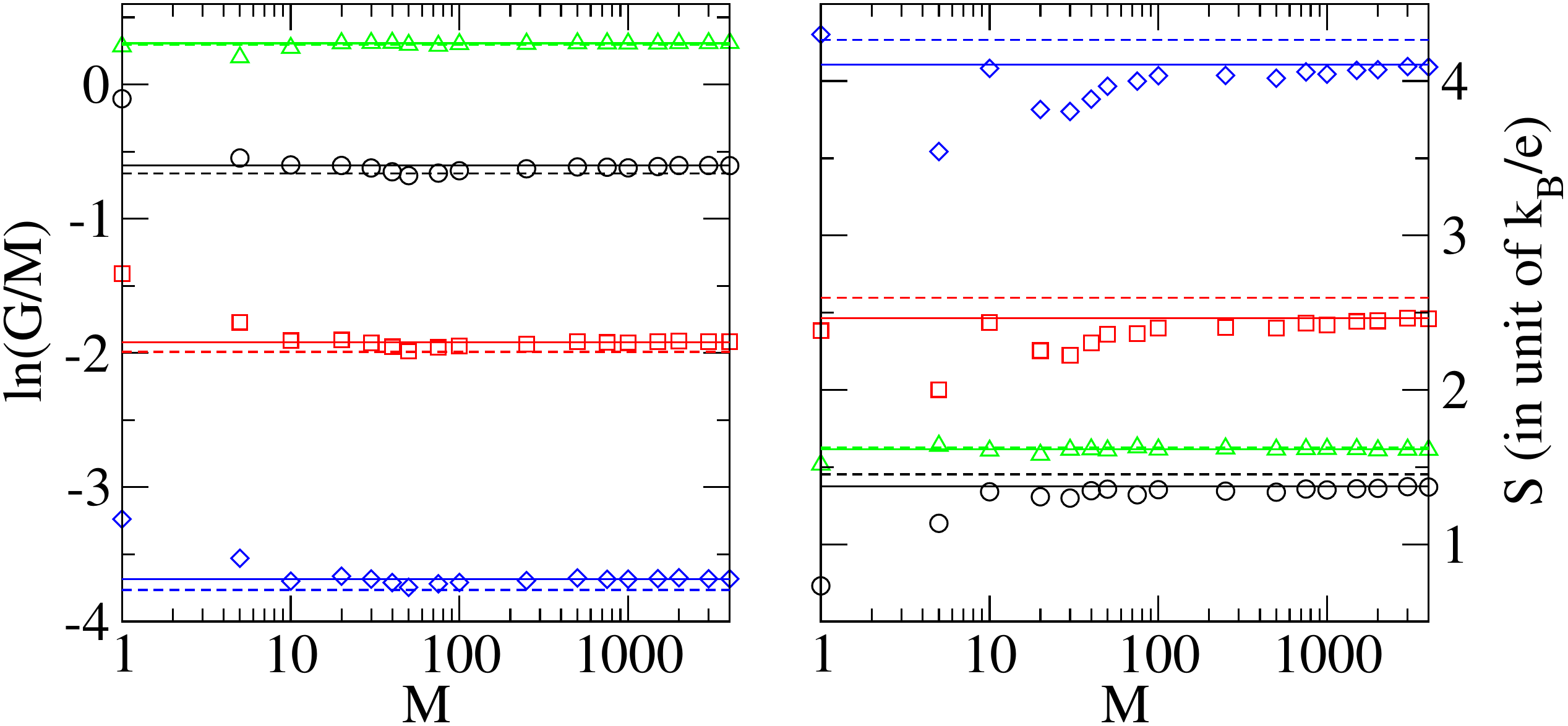}
\caption{Convergence of $G/M$ (left, in units of $e^2/\hbar$) and $S$ (right) with the number $M$ of parallel NWs. Symbols correspond to 
$V_g=1.9\,t$ ({\large$\circ$}), $2.1\,t$ ({\scriptsize{\color{red}$\square$}}) and $2.3\,t$ ({\large{\color{blue}$\diamond$}}) at $k_BT=0.1\,t$, and 
$V_g=1.9\,t$ at $k_BT=0.5\,t$ ({\scriptsize{\color{green}$\triangle$}}). The horizontal lines indicate the corresponding mean values (dashed lines) 
and typical values (solid lines) of $\ln G$ and $S$ of a single wire ($M=1$). Parameters: $W=t$, $\gamma_e=\gamma_{ep}=t/\hbar$ and $L=450a$.}
\label{fig:fluctuations}
\end{figure}

The typical conductance $G_0$ and thermopower $S_0$ of a single NW were studied in Ref.~\cite{Bosisio20142}.  They are defined as the {\it median} of 
the distribution of $\ln G$ and $S$, obtained when considering a large statistical ensemble of disorder configurations.  In Fig.~\ref{fig:fluctuations} 
we show that, if the system is made of a sufficiently large number $M$ of parallel NWs, the \emph{overall} electrical conductance scales as the number 
of wires times the typical NW value ($G\approx M\, G_0$), while the thermopower averages out to the typical value of a single wire ($S\approx S_0$). 
For completeness the {\it mean} values are also shown and seen to be a less accurate estimate.  As expected, convergence is faster at higher temperatures. 
Identical results have been obtained for the electronic thermal conductance $K^e\approx MK^e_0$ (not shown).
\section{Power factor and figure of merit}
\label{sec_QZT}
\begin{figure}
    \includegraphics[keepaspectratio,width=0.9\columnwidth]{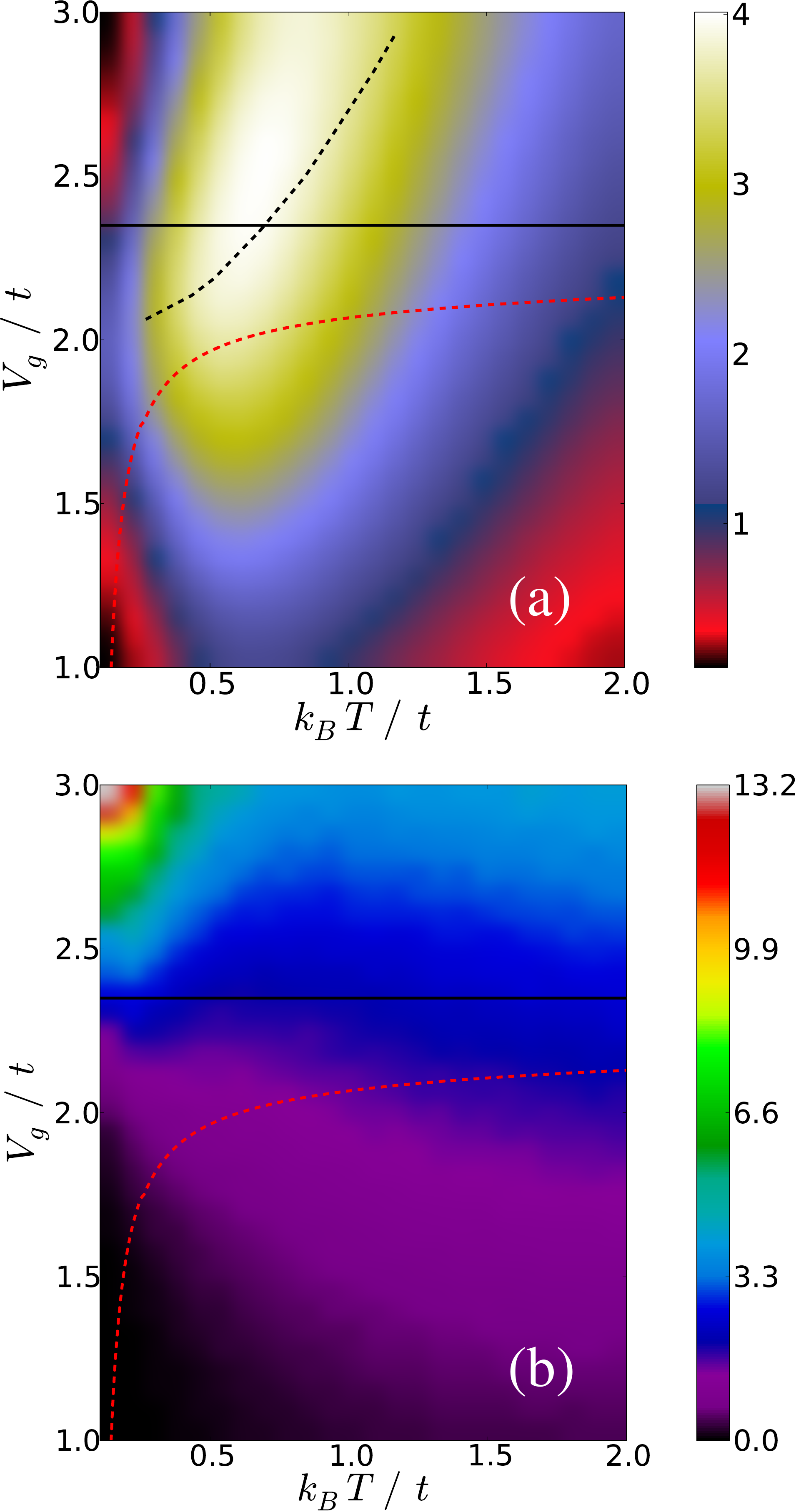}   
    \caption{(Color online) $\mathcal{Q}/M$ in units of $k_B^2/\hbar$ (a) and $Z_eT$~(b) as a function of $T$ and $V_g$. Data are shown in the 
large~$M$ limit ($M=150$) where there is self-averaging. The horizontal lines give $V_g$'s value at which the band edge is probed at 
$\mu$ (below [above] it, one probes the inside [outside] of the impurity band). The red dashed lines $T=T_M$ separate the VRH ($T\lesssim T_M$) 
and the NNH ($T\gtrsim T_M$) regimes. The black dashed line in (a) is the contour along which $S_0=2k_B/e$. Parameters: 
$W=t$, $\gamma_e=\gamma_{ep}=t/\hbar$ and $L=450a$.}%
   \label{fig:QZT}
\end{figure}

By stacking a large number $M$ of NWs in parallel, the device power factor can be enhanced  
$\mathcal{Q}\approx MS_0^2G_0$ {\it without} affecting its electronic figure of merit $Z_eT\approx S_0^2G_0T/K_0^e$. Fig.~\ref{fig:QZT} shows 
how the asymptotical $\mathcal{Q}/M$ and $Z_eT$ values (reached when $M\gtrsim 100$) depend on the gate voltage $V_g$ and on the temperature $T$. 
We observe in panel~(a) that the power factor is maximum for $\mu$ close to the impurity band edge (black solid line) and for VRH temperatures.  
This parameter range represents the best compromise between two opposite requirements: maximizing the thermopower (hence favoring low $T$ 
and large $V_g$) while keeping a reasonable electrical conductance (favoring instead higher $T$ and $V_g\approx 0$). Formulas previously 
reported,\cite{Bosisio20142} giving the $T$- and $V_g$-dependence of $G_0$ and $S_0$, let us predict that $\mathcal{Q}$ is maximal when 
$|S_0|=2k_B/|e|\approx 0.2\,\mathrm{mV}\,\mathrm{K}^{-1}$ (black dashed line).  A comparison between panels (a) and (b) of Fig.~\ref{fig:QZT} 
reveals that, in the parameter range corresponding to the best power factor ($V_g\sim2.5t, k_BT\sim0.6t$), $Z_eT\simeq 3$, a remarkably large value.  
Much larger values of $Z_eT$ could be obtained at lower temperatures or far outside the band, but they are not of interest for practical 
purposes since in those regions $\mathcal{Q}$ is vanishing. 
In Appendix \ref{app_size}, $\mathcal{Q}$ and $Z_eT$ are shown to be roughly independent of the NWs length $L$ (for $L\gtrsim L_M$) in the 
temperature and gate voltage ranges explored in Fig.~\ref{fig:QZT}. Moreover $\mathcal{Q}/\gamma_e$ and $Z_eT$ are almost independent 
of the choice of the parameters $\gamma_e$ and $\gamma_{ep}$, provided $\gamma_{ep}\gtrsim\gamma_e$ (see Appendix~\ref{app:depgammaeep}). 
When $\gamma_{ep}<\gamma_e$, both quantities are found to be (slightly) reduced.\\
\indent Let us now estimate the order-of-magnitude of the device performance.  The substrate (or the NWs themselves if they are suspended) 
is assumed to supply enough phonons to the NWs charge carriers for the condition $\gamma_{ep}\gtrsim\gamma_e$ to hold.  Besides, we keep explicit 
the $\gamma_e$-linear dependence of $\mathcal{Q}$ (and of $K^e_0$ that will soon be needed). $\gamma_e$ depends on the quality of the metal/NW contact.  
We estimate it to be within the range $0.01-1$ in units of $t/\hbar$, where $t/k_B\approx 150\,\mathrm{K}$ throughout
\footnote{We estimated $t$ by comparing the band width $4t+W$ in our model to the typical width of the impurity band in highly doped Silicon NWs 
(see for instance Ref.~\cite{Salleh2011}). Note that the NWs are then depleted by field effect.}.
This yields $\gamma_e\approx 0.02-2\times 10^{13}\,\mathrm{s}^{-1}$. For the sake of brevity, we introduce the dimensionless number 
$\tilde{\gamma}_e=\gamma_e\hbar/t$. Focusing on the region of Fig.~\ref{fig:QZT}(a) where the power factor is maximal, we evaluate the typical 
output power and figure of merit than can be expected.  We first notice that power factor $\mathcal{Q}/M\approx4k_B^2/h$ maximum values in 
Fig.~\ref{fig:QZT}(a), obtained with $\tilde{\gamma}_e=1$, would yield $\mathcal{Q}\approx 7\tilde{\gamma}_e\times 10^{-7}\,\mathrm{W}.\mathrm{K}^{-2}$ 
for a chip with $M\approx 10^5$ parallel NWs. Since $\mathcal{Q}$ controls the maximal output power $P_{max}$ that can be extracted from 
the setup as $P_{max}=\mathcal{Q}(\delta T)^2/4$,\cite{Benenti2013} one expects $P_{max}\approx 20\tilde{\gamma}_e\,\mu\mathrm{W}$ for a small temperature 
bias $\delta T\approx 10\,\mathrm{K}$.  In this region a large value $Z_eT\approx 3$ is obtained, but to estimate the full figure of merit 
$ZT=Z_eT/(1+K^{ph}/K^{e})$, the phononic part $K^{ph}$ of the thermal conductance must also be taken into account.  To limit the reduction of $ZT$ 
by phonons, the setup configuration with suspended nanowires is preferable [Fig.~\ref{fig:sys}(a)].  In this case $K^{ph}\approx MK_0^{nw}$, $K_0^{nw}$ 
being the typical phononic thermal conductance of a single NW, and has to be compared to $K^{e}\approx MK_0^{e}$.  Introducing the corresponding 
conductivities $\kappa$'s, the ratio $K^{ph}/K^{e}\approx \kappa_0^{nw}/\kappa_0^{e}$ is to be estimated.  Our numerical results obtained for 1D NWs 
show $K_0^{e}\approx 1.5\tilde{\gamma}_e k_Bt/\hbar$ in the range of interest where $\mathcal{Q}$ is maximal and $Z_eT\approx 3$ (at $V_g=2.5t$ and 
$k_BT=0.6t$, keeping other parameters in Fig.~\ref{fig:QZT} unchanged). To deduce the corresponding conductivity $\kappa_0^{e}$, the NW aspect ratio must 
be specified. We consider for instance the case of $1\,\mu\mathrm{m}$-long NWs with a diameter of $20\,\mathrm{nm}$, for which our pure 1D model is 
expected to hold\footnote{The use of the 1D model is justified at a semi-quantitative level if the nanowire diameter is smaller than the Mott hopping 
length $L_M$ (the typical length of an electron hop along the nanowire).}, at least semi-quantitatively. Thereby we get 
$\kappa_0^{e}\approx 1\tilde{\gamma}_e\,\mathrm{W}/(\mathrm{K.m})$, while the measured thermal conductivity of Si NWs of similar 
geometry is $\kappa_0^{nw}\approx 2\,\mathrm{W}/(\mathrm{K.m})$ at $T\approx 100\,\mathrm{K}$.\cite{Li2003}  We thus evaluate for suspended NWs 
$ZT\approx Z_eT/(1+2/\tilde{\gamma}_e)$, i.e. $ZT\approx 0.01-1$ for $Z_eT\approx 3$ and $\tilde{\gamma}_e=0.01-1$. Those estimations though rough 
are extremely encouraging as they show us that such a simple and Si-based device shall generate high electrical power from wasted heat (scalable 
with $M$, for $M$ large enough) with a fair efficiency (independent of $M$, for $M$ large enough).\\
\indent Let us note that maximizing $\gamma_e$ is important for achieving high $\mathcal{Q}$ and $ZT$. However at the same time 
$\gamma_{ep}\gtrsim\gamma_e$ should preferably hold. If the NWs themselves do not ensure a large enough $\gamma_{ep}$, the use of a 
substrate providing phonons is to be envisaged. Yet, this will add a detrimental contribution $K^{sub}$ to $K^{ph}$. In general the 
substrate cross-section ($\Sigma^{sub}$) will be substantially larger than the NWs one ($M\Sigma^{nw}$). Thus, even for a good thermal 
insulator such as SiO$_2$, with thermal conductivity $\kappa^{sub}\approx 0.7\,\mathrm{W}/(\mathrm{K.m})$ at 
$T\approx 100\,\mathrm{K}$,\cite{Hung2012} $Z/Z_e=[1+(\kappa^{sub}\Sigma^{sub}+M\kappa_0^{nw}\Sigma^{nw})/M\kappa_0^{e}\Sigma^{nw}]^{-1}\ll1$. 
Better ratios $Z/Z_e$ could be obtained for substrates with lower $K^{sub}$ (Silica aerogels,\cite{Hopkins2011} porous silica,\cite{Scheuerpflug1992} 
very thin substrate layer) but they will not necessarily guarantee a good value of $\gamma_{ep}$ (and hence of $Z_e$). Clearly, finding a balance between 
a large $\gamma_{ep}$ and a low $K^{ph}$ is a material engineering optimization problem. Though the presence of a substrate appears detrimental for 
efficiently harvesting electrical energy from the wasted heat, we shall now see how it could be used for heat management at the nanoscale.
\begin{figure*}
    \includegraphics[keepaspectratio,width=0.9\textwidth]{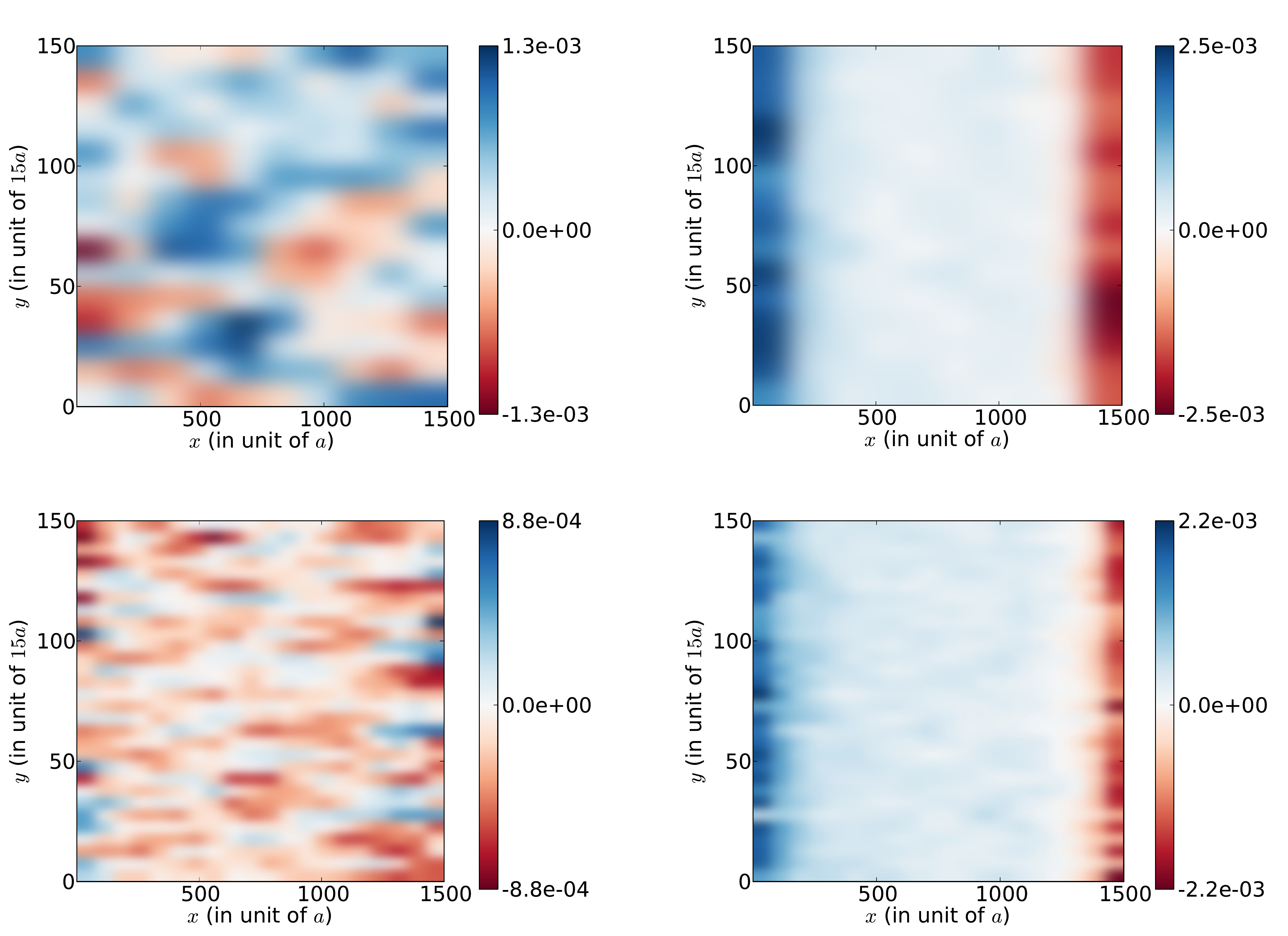}
    \caption{(Color online) Map of the local heat exchanges $\mathcal{I}^Q_{x,y}$ between the NWs and the phonon bath (substrate), in units of $10^{-3}t^2/\hbar$, 
at the band center ($V_g=0$, left) and near the lower band edge ($V_g=2.25t$, right), for $k_BT=0.25t$ (top) and $k_BT=0.5t$ (bottom). When phonons are 
absorbed by NWs charge carriers in the small area of size $\Lambda_{ph}^2$ around $(x,y)$, $\mathcal{I}^Q_{x,y}>0$ and the substrate below is locally cooled 
down (blue). When phonons are released, $\mathcal{I}^Q_{x,y}<0$ and the substrate is locally heated up (red). As explained in the text, we took 
$\Lambda_{ph}=75a$ for $k_BT=0.5t$ and $\Lambda_{ph}=150a$ for $k_BT=0.25t$. Note that the formation of hot and cold spots at the boundaries of the NWs is 
clearly visible for both temperatures when $V_g$ is tuned in order to probe their band edges (right), while no net effect is evident in absence of any 
gate voltage (left). In all panels, data have been plotted for $M=150$ NWs of length $L=1500a$ with interspacing $15a$. Other parameters are $W=t$, 
$\gamma_{e}=\gamma_{ep}=t/\hbar$ and $\delta\mu=10^{-3}t$.}%
   \label{fig:hotspots}
\end{figure*}
\section{Gate-controlled creation/annihilation of cold/hot strips}
\label{sec_hotspots}
Hereafter, we consider the deposited setup sketched in Fig.~\ref{fig:sys}(b) and assume a constant temperature 
$T$ everywhere.  An intriguing feature of this three-terminal setup is the possibility to generate/control hot/cold strips close to the substrate 
boundaries by applying a bias $\delta\mu/e$, if one tunes $V_g$ for probing the NWS band edges.  This effect is a direct consequence of the heat 
exchange mechanism between electrons in the NWs and phonons in the substrate.  Indeed, given a pair of localized states $i$ and $j$ inside a NW, 
with energies $E_i$ and $E_j$ respectively, the heat current absorbed from (or released to) the phonon bath by an electron in the transition 
$i\to j$ is $I_{ij}^{Q}=\left(E_j-E_i\right)I^N_{ij}$, $I^N_{ij}$ being the hopping particle current between $i$ and $j$.\cite{Bosisio20142} 
The overall hopping heat current through each localized state $i$ is then found by summing over all but the $i$-th states:
\be\label{eq:IQ_i}
I_{i}^{Q}=\sum_j I_{ij}^Q = \sum_{j} \left(E_j-E_i\right)I^N_{ij}
\ee
with the convention that $I^Q_{i}$ is positive (negative) when it enters (leaves) the NWs at site $i$.  
Since the energy levels $E_i$ are randomly distributed, the $I^Q_i$'s (and in particular their sign) fluctuate from site to site 
(see Fig.~\ref{fig:HotSpots_Vg2.25_kT0.05} in Appendix~\ref{app_lambdaph} for an illustration). The physically relevant quantities are 
however not the $I^Q_i$'s, rather their sum within an area $\Lambda_{ph}\times\Lambda_{ph}$, where $\Lambda_{ph}$ is the phonon thermalization 
length in the substrate (i.e. the length over which a {\it local} substrate temperature can be defined, see Appendix~\ref{app_lambdaph} for an estimation). 
Given a point $(x,y)$ and a $\Lambda_{ph}\times\Lambda_{ph}$ area centered around it, such sum is denoted $\mathcal{I}^Q_{x,y}$.  
If $\mathcal{I}^Q_{x,y}>0$ $[<0]$, a volume $\Lambda_{ph}^3$ of the substrate beneath $(x,y)$ is cooled [heated] \footnote{Practically, we map the 2D 
parallel NW array onto a square grid, and for each square of size $\Lambda_{ph}^2$ we calculate the net heat current entering the NWs. For better 
visibility, data are then smoothed (with a standard Gaussian interpolation) to produce the heat map shown in Fig.~\ref{fig:hotspots}.}. Deeper 
than $\Lambda_{ph}$ away from the surface, the equilibrium temperature $T$ is reached.\\
\indent Fig.~\ref{fig:hotspots} shows how $\mathcal{I}^Q_{x,y}$ depends on the coordinates $x, y$ in the two-dimensional parallel NW array. 
Left and right panels show respectively the situation in the absence of a gate voltage, when charge carriers tunnel into/out of NWs at 
the impurity band \textit{center}, and the opposite situation when a large gate voltage is applied in order to inject/extract carriers at 
the band \textit{bottom}. In both cases, two values of the temperature are considered (top/bottom panels). All other parameters are fixed. 
Note that data are plotted for the model introduced in Sec.~\ref{sec_method}, having estimated $a\approx 3.2\,\mathrm{nm}$, 
$t/k_B\approx 150\,\mathrm{K}$, and $\Lambda_{ph}\approx 480[240]\,\mathrm{nm}\approx 150[75]a$ for SiO$_2$ substrate at the temperatures considered, 
$T=0.25[0.5]t/k_B\approx37.5[75]\,\mathrm{K}$. Those estimates are discussed in Appendix \ref{app_lambdaph}. In the left panels of Fig.~\ref{fig:hotspots}, 
the heat maps show puddles of positive and negative $\mathcal{I}^Q_{x,y}$, corresponding respectively to cooled and heated regions in the substrate below 
the NW array. They are the signature of random absorption and emission of substrate phonons by the charge carriers, all along their propagation through 
the NWs around the band center. In the right panels, the regions of positive and negative $\mathcal{I}^Q_{x,y}$ are respectively confined to the NWs 
entrance and exit. This is due to the fact that charge carriers entering the NWs at $\mu$ around the band bottom find available states to jump to 
(at a maximal distance $L_M$ in space or $\Delta$ in energy~\footnote{Here $L_M\approx 10.6a$ and $\Delta\approx 2.4t$ for $k_BT=0.25t$, while 
$L_M\approx 7.5a$ and $\Delta\approx 3.3t$ for $k_BT=0.5t$.}) only \textit{above} $\mu$. Therefore, they need to absorb phonons to reach higher energies 
states (blue region).  After a few hops, having climbed at higher energies, they continue propagating with equal probabilities of having upward/downward 
energy hops (white region). On reaching the other end they progressively climb down, i.e. release heat to the substrate (red region), until they reach $\mu$ 
and tunnel out into the right reservoir. As a consequence, the substrate regions below the NWs extremities are cooled on the source side and heated 
on the drain side [see Fig.~\ref{fig:sys}(b)]. A comparison between top and bottom panels of Fig.~\ref{fig:hotspots} shows us that the heat maps are not 
much modified when the temperature is doubled [from $k_BT=0.25t$ (top) to $k_BT=0.5t$ (bottom)]. The fact that the surface $\Lambda_{ph}\times\Lambda_{ph}$ 
inside which the heat currents are summed up is smaller at larger temperature ($\Lambda_{ph}=75a$ at $k_BT=0.5t$ instead of $\Lambda_{ph}=150a$ at $k_BT=0.25t$) 
is compensated by a smoothing of the $I^Q_i$'s fluctuations. This makes the hot and cold strips still clearly visible and well-defined in the bottom right 
panel of Fig.~\ref{fig:hotspots}.\\
\indent We point out that the maximum values of $\mathcal{I}^Q_{x,y}$ are roughly of the same order of magnitude with or without the gate (see scale 
bars in Fig.~\ref{fig:hotspots}).  The advantage of using a gate is the ability to split the positive and negative $\mathcal{I}^Q_{x,y}$ regions 
into two well separated strips in the vicinity of the injection and drain electrodes.  One can then imagine to exploit the cold strip in the substrate 
to cool down a hot part of an electronic circuit put in close proximity. Let us also stress that the assumption of elastic tunneling processes between 
the electrodes and the NWs is not necessary to observe the gate-induced hot/cold strips. The latter arise from the ``climbing'' up/down in energy that 
charge carriers, at $\mu$ far into the electrodes, must undergo in order to hop through the NWs (hopping transport being favored around the impurity 
band center in the NWs). Though in our model heat exchanges take place only inside the NWs, phonon emission/absorption will actually take place also at 
the electrodes extremities, roughly within an inelastic relaxation length from the contacts. This has clearly no qualitative impact, as it only amounts 
to a slight shift/smearing of the hot/cold strips. \\
\indent Finally, let us estimate the cooling powers associated to the data shown in Fig.~\ref{fig:hotspots}. Assuming again $t/k_B\approx 150\,\mathrm{K}$ 
and $a\approx 3.2\,\mathrm{nm}$, we find that a value of $\mathcal{I}^Q_{x,y}=10^{-3}(t^2/\hbar)$ in Fig.~\ref{fig:hotspots}(bottom) corresponds to a cooling 
power density of the order of $8.10^{-10}\,\mathrm{W}.\mu\mathrm{m}^{-2}$ at the temperature considered $T=0.5t/k_B\approx 77\,\mathrm{K}$ (the boiling 
temperature of liquid nitrogen at atmospheric pressure), for which $\Lambda_{ph}\approx 240\,\mathrm{nm}$ in SiO$_2$. We underline that this order of magnitude 
is obtained for a given set of parameters, in particular for an infinitesimal bias $\delta\mu=10^{-3}t \approx 13\, \mu V$ that guarantees to remain in the 
linear response regime. It should not be taken in the strict sense but only as a benchmark value to fix ideas. For instance, according to this estimation, one 
should be able to reach cooling power densities $\approx 6.10^{-8}\,\mathrm{W}.\mu\mathrm{m}^{-2}$ by applying a larger bias 
$\delta\mu/e \approx 1\,\mathrm{mV}$. To be more specific, we note that the geometry considered in Fig.~\ref{fig:hotspots} is realized\footnote{Data shown 
in Fig.~\ref{fig:hotspots} result from numerical simulations run for a set of 150 1D NWs (of length $1500a$) separated from each other by a distance $15a$. 
They are expected to describe the physics of realistic arrays made of 150 NWs covering an area of width $150\times 15a\approx 7.2\mu\mathrm{m}$ and length 
$1500a\approx 5\mu\mathrm{m}$, taking again $a\approx 3.2\,\mathrm{nm}$. For instance, 150 NWs with $10\,\mathrm{nm}$ diameter and $20\%$ packing density. 
Other configurations could be considered as well, as long as the NW diameter is small enough for the 1D model to make sense and the packing density does 
not exceed the typical values reachable experimentally.} with a bidimensional array of 150, $5\,\mu$m-long NWs, covered by two $7.2\,\mu$m-long (or longer) 
metallic electrodes. For this geometry and at $T\approx 77\,\mathrm{K}$, the areas of the cooled and heated regions are approximately $7.2 \times 0.25 \approx 2 
\mu\mathrm{m}^{2}$ (see the lower right panel of Fig.~\ref{fig:hotspots}) but if one considered $1\,\mathrm{cm}$ electrodes covering $2.10^{5}$ NWs, those areas 
would naturally extend. Thus, for a bias $\delta\mu/e \approx 1\,\mathrm{mV}$ and a temperature $T \approx 77 \,\mathrm{K}$, our setup would allow to take 
$\approx 0.15\,\mathrm{mW}$ in a strip of $1\,\mathrm{cm}\times 0.25\,\mu\mathrm{m}$ area and $0.25\,\mu\mathrm{m}$ thickness located in the SiO$_2$ substrate 
below the source electrode and to transfer it in another strip of similar size located at $5\,\mu\mathrm{m}$ away below the drain electrode. Obviously, the 
longer the NWs, the longer would be the scale of the heat transfer. The larger the bias and the number of used NWs, the larger would be the heat transfer. 
\section{Conclusion}
The low carrier density of a doped semiconductor can be varied by applying a voltage on a (back, side or front) metallic gate. This led us 
to study thin and weakly doped semiconductor NWs, where electron transport is activated, instead of thick metallic NWs (with much larger electrical and 
thermal conductances) where the field effects are negligible. Considering arrays of these NWs in the FET configuration, we have focused our attention 
on the activated regime which characterizes a very broad temperature domain in amorphous semiconductors.\cite{Ambegaokar1971} When charge transport between localized states 
is thermally assisted by phonons, we have shown that the absorption or the emission of phonons in strips located near the source and drain electrodes 
can be controlled with a back gate. This opens new perspectives for managing heat at submicron scales. By tuning the electrochemical potential $\mu$ near 
the band edges of the NWs impurity band, we have studied how to take advantage of electron-phonon coupling for energy harvesting and hot spot cooling. 
Our estimates indicate that large power factors are reachable in these arrays, with good thermoelectric figures of merit. 
\acknowledgments
This work was supported by CEA within the DSM-Energy Program (project E112-7-Meso-Therm-DSM). We thank O. Bourgeois, Y. Imry and F. Ladieu for stimulating 
discussions.
\appendix
\section{Resolution of the Random Resistor Network problem}
\label{app_rrn}
\begin{figure}
    \begin{center}
      \includegraphics[keepaspectratio, width=\columnwidth]{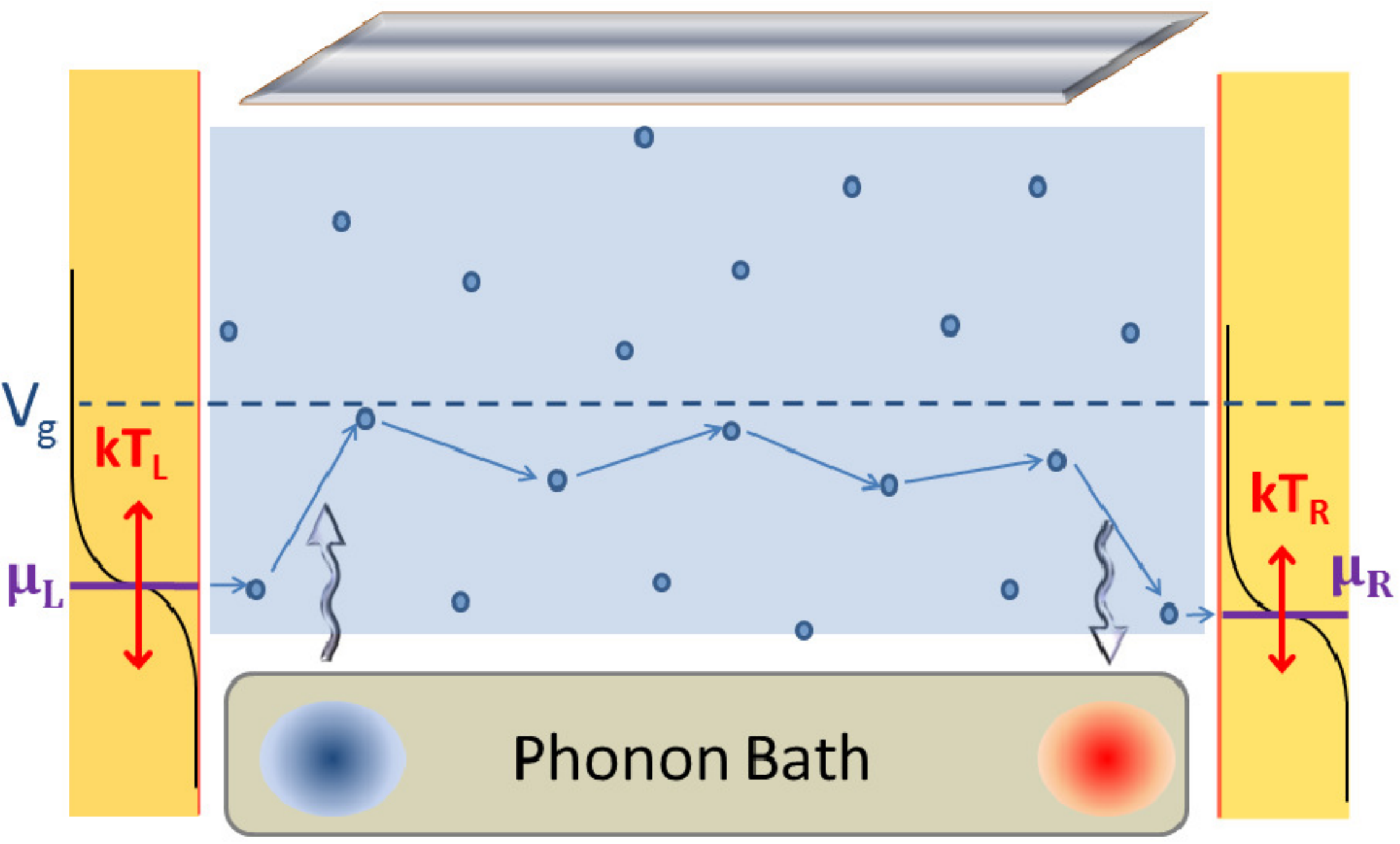}
    \end{center}
    \caption{(Color online) Phonon-assisted hopping transport through the localized states (dots) of a disordered NW connected to two electrodes $L$ and 
$R$, and to a phonon bath. The electronic reservoirs $L$ and $R$ are thermalized at temperatures $T_{L[R]}$ and held at electrochemical potentials $\mu_{L[R]}$ 
(their Fermi functions are sketched by the black curves on both sides). A metallic gate (shaded grey plate drawn on top) allows to shift the NW impurity band 
(blue central region). Here, the gate potential $V_g$ is adjusted such that electrons tunnel in and out of the electronic reservoirs near the lower edge of 
the impurity band. Therefore, electrons tend to absorb phonons at the entrance in order to reach available states of higher energies, and to emit phonons 
on the way out. The two wavy arrows indicate the local heat flows between the NW electrons and the phonon bath. They give rise to a pair of cold (blue) and 
hot (red) spots in the substrate beneath the NW (in the deposited setup configuration).}
   \label{fig:sketch_sys}
\end{figure}

Hereafter, we summarize the numerical method used to solve the random-resistor network problem.\cite{Miller1960,Ambegaokar1971,Jiang2013} 
The three-terminal setup configuration is reminded in Fig.~\ref{fig:sketch_sys}, with emphasis on the hopping transport mechanism taking place in the NWs. 
Starting from a set of states $i$ localized at positions $x_i$ inside the NWs, with energies $E_i$ and localization lengths $\xi_i$, we first evaluate 
the transition rates $\Gamma_{i\alpha}$ from the localized state $i$ to the reservoir $\alpha=L$ or $R$, and $\Gamma_{ij}$ from states $i$ to $j$ within 
the same wire ({\it inter}-wire hopping being neglected). They are given by the Fermi Golden rule as
\begin{align}
\Gamma_{i\alpha}&=\gamma_{i\alpha} f_i [1-f_{\alpha}(E_i)]\label{eq_rates1}\\
\Gamma_{ij}&=\gamma_{ij} f_i (1-f_j) [N_{ij}+\theta(E_i-E_j)]\label{eq_rates2}
\end{align}
where $f_i$ is the occupation probability of state $i$, $f_\alpha(E)=[\exp((E-\mu_\alpha)/k_BT_\alpha)+1]^{-1}$ is the Fermi distribution of reservoir 
$\alpha$, $N_{ij}=[\exp(|E_j-E_i|/k_BT)-1]^{-1}$ is the probability of having a phonon with energy $|E_j-E_i|$ assisting the hop, and $\theta$ is the 
Heaviside function. In Eq.~\eqref{eq_rates1}, $\gamma_{i\alpha}=\gamma_e\exp(-2x_{i\alpha}/\xi_i)$, $x_{i\alpha}$ denoting the distance of state $i$ 
from reservoir $\alpha$, and $\gamma_e$ being a constant quantifying the coupling from the localized states in the NW to the extended states in the 
reservoirs. Usually, $\xi_i\approx\xi(\mu)$ is assumed and the rate $\gamma_{ij}$ in Eq.~\eqref{eq_rates2} is simply given by 
$\gamma_{ij}=\gamma_{ep}\exp(-2x_{ij}/\xi(\mu))$, with $x_{ij}=|x_i-x_j|$ and $\gamma_{ep}$ measuring the electron-phonon coupling. Since this approximation 
does not hold in the vicinity of the impurity band edges, where the localization lengths vary strongly with the energy, we use a generalized expression 
for $\gamma_{ij}$ that accounts for the different localization lengths $\xi_i\neq\xi_j$ (see Ref.~\cite{Bosisio20142}).\\
\indent By using Eqs.~\eqref{eq_rates1}-\eqref{eq_rates2} and imposing charge conservation at each network node $i$, we deduce the $N\,f_i$'s of the 
$M$ independent NWs. The charge and heat currents flowing from reservoir $\alpha$ to the system can then be calculated as $I^e_{\alpha}=e\,\sum_i I_{\alpha i}$ 
and $I^Q_{\alpha}=\sum_i I_{\alpha i} (E_i-\mu_{\alpha})$, where $I_{\alpha i}=\Gamma_{\alpha i}-\Gamma_{i\alpha}$ and $e$ is the electron charge. In principle, 
the heat current $I^Q_P=(1/2)\sum_i I^Q_i$ coming from the phonon bath can be calculated as well but in this work, we only investigated the behavior of 
the local heat currents $I_{i}^{Q}= \sum_{j} \left(E_j-E_i\right)I^N_{ij}$ with $I^N_{ij}=\Gamma_{ij}-\Gamma_{ji}$. Without loss of generality, we choose 
the right terminal $R$ as the reference, i.e. we set $\mu_R=\mu$, $T_R=T$ and we impose on the left side $\mu_L=\mu+\delta\mu$, $T_L=T+\delta T$. Using 
the Onsager formalism, we relate the particle ($I^e_{L}$) and heat ($I^Q_{L}$) currents computed in linear response to the small imposed bias $\delta\mu$ 
and $\delta T$.\cite{Callen1985} This allow us to deduce the thermoelectric coefficients $G$, $K^e$ and $S$.

\section{Size Effects}
\label{app_size}
\begin{figure}
\centering
\includegraphics[keepaspectratio, width=\columnwidth]{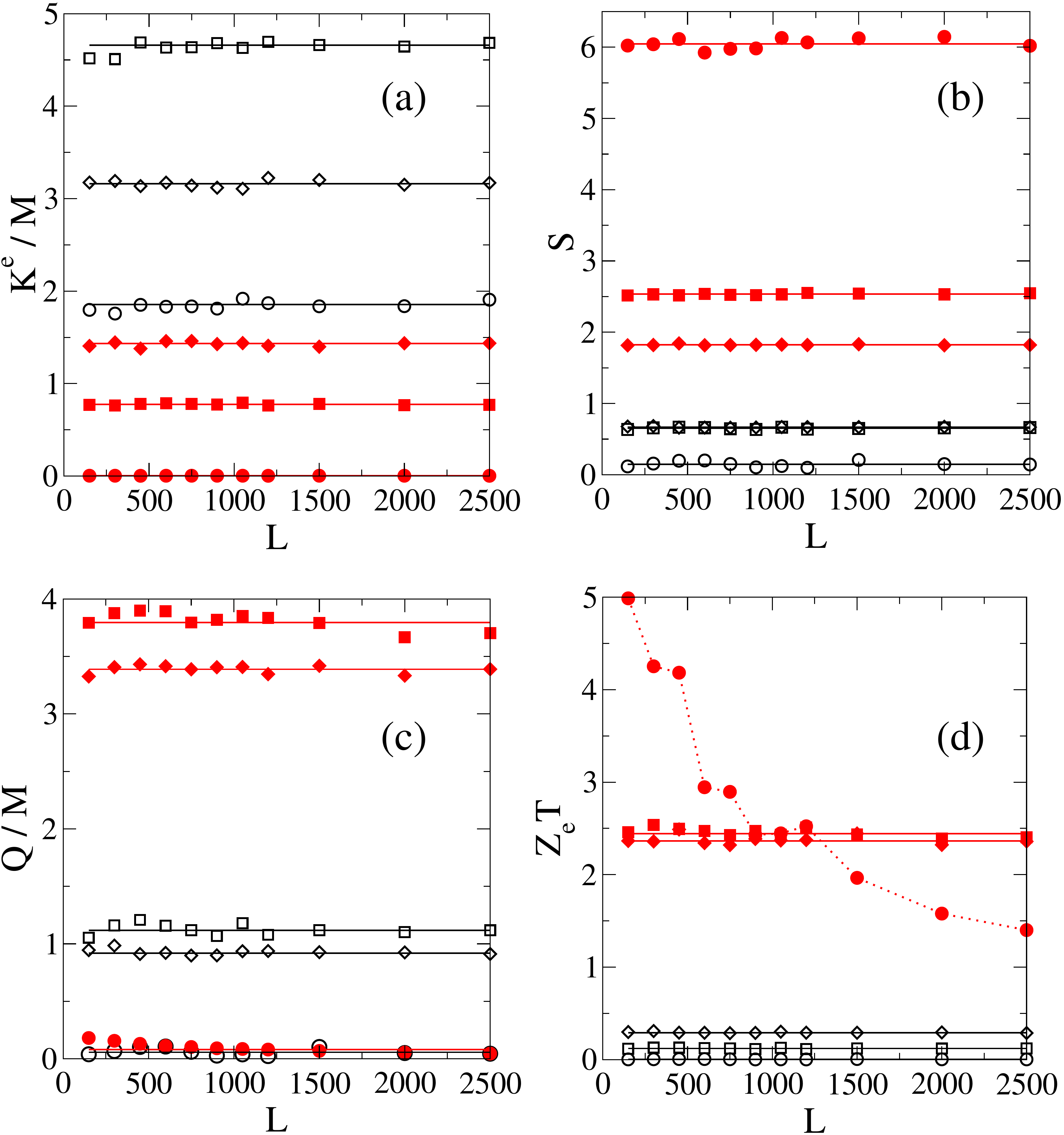}
\caption{(Color online) Behavior of the transport coefficients as a function of the NWs length $L$ (in units of the spacing $a$). Panels show (a) 
the rescaled electronic contribution to the thermal conductance $K^e$ (in units of $k_Bt/\hbar$), (b) the thermopower $S$ (in units of $k_B/e$), 
(c) the rescaled power factor $\mathcal{Q}$ (in units of $k_B^2/\hbar$) and (d) the electronic figure of merit $Z_eT$. In all the four panels, 
data are plotted for $k_BT=0.1t$ (circles), $k_BT=0.5t$ (squares) and $k_BT=t$ (rhombus), in the case of bulk transport ($V_g=t$, empty symbols) 
and edge transport ($V_g=2.5t$, full symbols). Lines are guides to the eye. Other parameters are fixed to $W=t$ and $\gamma_e=\gamma_{ep}=t/\hbar$.}
\label{fig:sizeeffects}
\end{figure}

We have investigated the effects on the various transport coefficients $G$, $K^e$ and $S$, the power factor $\mathcal{Q}$, and the electronic figure 
of merit $Z_eT$, of varying the length $L$ of the NWs. The results are shown in Fig.~\ref{fig:sizeeffects}, for three values of the temperatures 
$k_BT=0.1t,0.5t$ and $1.0t$, and for two configurations corresponding to bulk ($V_g=t$) and edge transport ($V_g=2.5t$). In all cases (except the one 
for $k_BT=0.1t$ and $V_g=2.5t$, $\text{\small{\color{red}$\bullet$}}$ in Fig.~\ref{fig:sizeeffects}), electronic transport through the NWs is thermally 
activated (see Fig.3 in Ref.~\cite{Bosisio20142}) and the results are seen to be essentially size-independent, as expected in the activated regime. 
In the case identified by $\text{\small{\color{red}$\bullet$}}$ in Fig.~\ref{fig:sizeeffects} and corresponding to the lowest temperature and the 
vicinity of the band edge, transport turns out to be achieved by elastic tunneling processes: the electrical conductance becomes size-dependent, 
which causes the electronic figure of merit $Z_eT$ to decrease roughly as $1/L$. However, being interested in the activated regime and in particular 
in the regime of temperatures where the power factor is largest ($k_BT\simeq 0.5t$), we can conclude that the size effects on the results shown in 
this work are completely negligible. Also, we note that the small fluctuations observed especially at the smallest sizes in Fig.~\ref{fig:sizeeffects} 
are a consequence of having taken a finite number of parallel NWs (M=150): they would diminish in the limit M$\to\infty$ due to self-averaging.
\section{On the dependence on the couplings $\gamma_e$ and $\gamma_{ep}$}
\label{app:depgammaeep}
\begin{figure*}
\centering
\includegraphics[keepaspectratio, width=\textwidth]{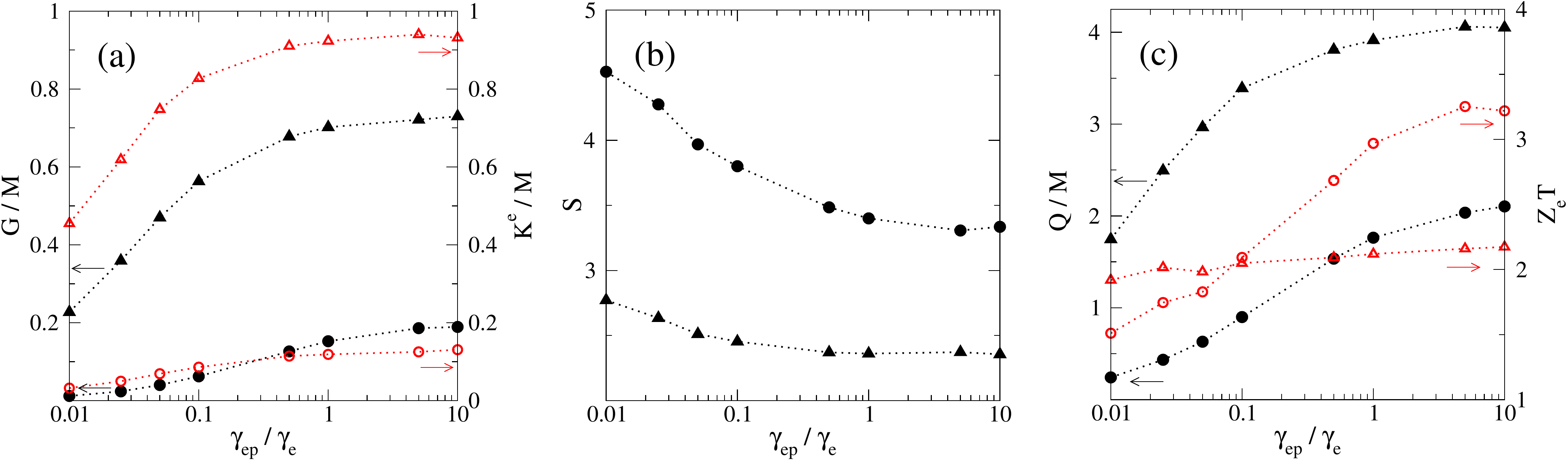}
\caption{(Color online) Dependency of $G$, $K^e$, $S$, $\mathcal{Q}$ and $Z_eT$ on the ratio $\gamma_{ep}/\gamma_e$. (a) Electrical 
($G/M$, black full symbols) and thermal ($K^e/M$, red empty symbols) conductances, in units of $e^2/\hbar$ and $k_B t/\hbar$ respectively. 
(b) Thermopower in units of $k_B/e$. (c) $\mathcal{Q}/M$ in units of $k_B^2/\hbar$ (black full symbols) and $Z_eT$ (red empty symbols). 
In all panels, different symbols correspond to $k_BT=0.2t$ (circles) and $k_BT=0.5t$ (triangles), while dotted lines are guides to the eye. 
Data have been plotted for a given set of $M=150$ parallel NWs of length $L=450a$, with $\gamma_e=t/\hbar$, $W=t$ and $V_g=2.4t$. Note that 
when $\gamma_{ep}\gtrsim\gamma_{e}$ all these coefficients are nearly constant.}
\label{fig:QZT_game_gamep_full}
\end{figure*}

In this section, we investigate how the transport coefficients $G$, $K^e$ and $S$, the power factor $\mathcal{Q}=S^2G$ and the electronic figure 
of merit $Z_eT=S^2GT/K^e$ are modified upon varying the couplings $\gamma_e$ and $\gamma_{ep}$ of the localized states with the electrodes and 
the phonon bath, respectively. We introduce the notation $\alpha\equiv \gamma_{ep}/\gamma_{e}$. We first notice that if $\alpha$ is kept fixed, 
the electrical conductance $G$ and the electronic thermal conductance $K^e$ are strictly proportional to $\gamma_e$, while the thermopower $S$ 
is independent of it. This behavior is a direct consequence of the formulation of the random resistor network problem and can be seen at the 
stage of writing the equations (see Ref.~\cite{Bosisio20142}), before solving them numerically. Therefore, for any fixed $\alpha$, 
$\mathcal{Q}/\gamma_e$ and $Z_eT$ are necessarily independent of the choice of $\gamma_e$. We thus find that $G/\gamma_{e}$, $K^e/\gamma_{e}$, 
$S$, $\mathcal{Q}/\gamma_e$ and $Z_eT$ are functions of the single parameter $\alpha$, and not of the couple of parameters $\gamma_e$ and 
$\gamma_{ep}$ separately. Those functions are plotted in Fig.~\ref{fig:QZT_game_gamep_full} for two different temperatures. The conductances, 
the power factor and the figure of merit increase with $\alpha$ (as long as lack of phonons is a limiting factor to transport through the NWs), 
while the thermopower decreases. All of them tend to saturate for $\alpha\gtrsim 1$. This shows us, \emph{inter alia}, that $\mathcal{Q}/\gamma_e$ 
and $Z_eT$ are essentially independent of $\gamma_e$ and $\gamma_{ep}$ if $\gamma_{ep}\gtrsim \gamma_e$ and that they only deviate slowly from 
this limit if $\gamma_{ep}< \gamma_e$. Such a robustness of $\mathcal{Q}/\gamma_e$ and $Z_eT$ to variations of $\gamma_e$ and $\gamma_{ep}$ 
reinforces the impact of the results shown in this work.
\section{Estimation of the phonon thermalization length}
\label{app_lambdaph}
\begin{figure}
\centering
\includegraphics[keepaspectratio, width=0.8\columnwidth]{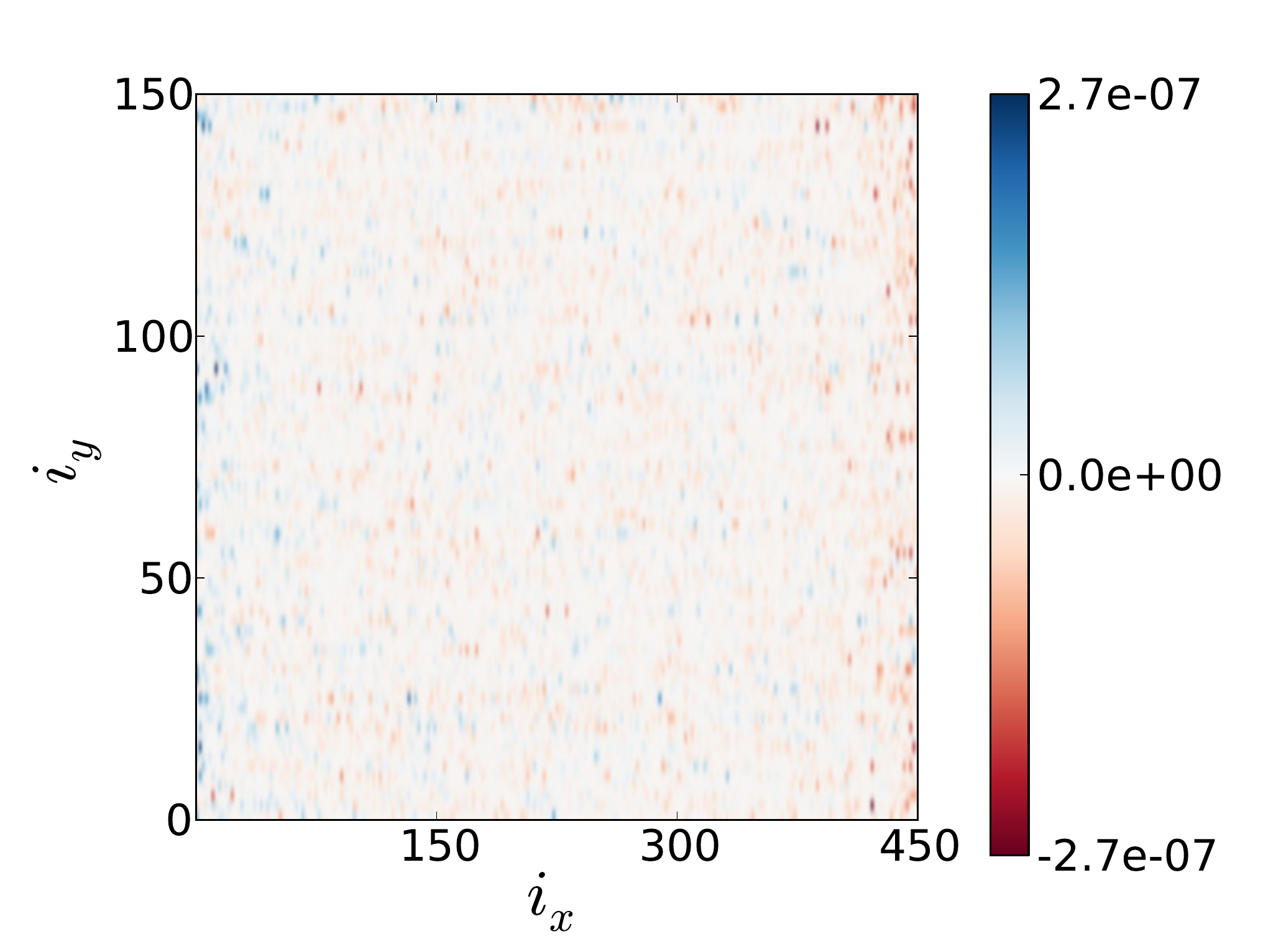}
\caption{(Color online) Map of the local heat currents $I^Q_i$ exchanged between the NWs and the substrate at each NWs site $i=(i_x,i_y)$, in units of $10^{-7}t^2/\hbar$, 
for $k_BT=0.05t$ and $V_g=2.25t$. The horizontal coordinate is the position along the NWs, while the vertical one labels each NW. The 
presence of hot and cold strips is hidden by the fluctuations. They emerge when the raw $I^Q_i$ data are summed up within areas of size 
$\Lambda_{ph}\times \Lambda_{ph}$. Parameters: $M=150$, $L=450a$, $W=t$, $\gamma_{e}=\gamma_{ep}=t/\hbar$ and $\delta\mu=10^{-5}t$.}
\label{fig:HotSpots_Vg2.25_kT0.05}
\end{figure}

We show in Fig.~\ref{fig:HotSpots_Vg2.25_kT0.05} an example of the map of the \emph{raw} heat currents $I^Q_i$ locally exchanged between 
the NWs and the substrate [see Eq.~\eqref{eq:IQ_i}]. We see that the $I^Q_i$'s fluctuate between positive and negative values at random 
positions of the substrate, and that no net effect emerges. As discussed in Sec.~\ref{sec_hotspots}, the formation of the hot and cold strips 
is a process which becomes visible only upon summing in a single term $\mathcal{I}^Q_{x,y}$ all the contributions $I^Q_i$ coming from states 
$i$ located within an area $\Lambda_{ph}\times\Lambda_{ph}$ around the point of coordinates $(x,y)$. $\Lambda_{ph}$, which represents the 
thermalization length of the substrate, is given by the inelastic phonon mean free path: this quantity may be different for different phonon 
wavelengths, and while it does not change much around room temperatures, it can vary significantly at lower temperatures. It is possible to 
relate $\Lambda_{ph}$ to the \emph{dominant} phonon wave length \cite{Pohl2002} as $\Lambda_{ph}=300 \lambda_{ph}^{dom}$, where the coefficient 
300 is for SiO$_2$ and may be different for other materials. This allows the calculation of the thermalization length $\Lambda_{ph}$, once 
$\lambda_{ph}^{dom}$ is known. According to Refs. \cite{Klitsner1987,Ziman1996}, the latter can be estimated as 
\begin{equation}
\lambda_{ph}^{dom}\simeq\frac{h v_s }{4.25 k_BT},
\label{eq:lambdaph}
\end{equation}
where $h$ is the Planck constant. Taking $v_s=5300\,\mathrm{m/s}$ the sound velocity in SiO$_2$,\cite{Ziman1996} we can easily deduce 
$\lambda_{ph}^{dom}\simeq 0.2\,\mathrm{nm}$ from which $\Lambda_{ph}\simeq 60\,\mathrm{nm}$ at room temperature $T=300\,\mathrm{K}$. 
Values of $\Lambda_{ph}$ at other (not vanishing) temperatures follow immediately from the temperature dependence in Eq.~\eqref{eq:lambdaph}. 
We shall stress that the real values of $\Lambda_{ph}$ may differ from our prediction by a small numerical factor, which however is not 
important within our qualitative approach. To convert these lengths in the units used in Sec. \ref{sec_hotspots}, we assume the average 
distance between localized states $a\approx 3.2\,\mathrm{nm}$ in highly doped silicon NWs, which together with  $t/k_B\approx 150 K$ allows 
us to estimate for example $\Lambda_{ph}\approx 75 a$ at $T=0.5t/k_B=75\,\mathrm{K}$.

\bibliography{PRAppl_BGFP}
\end{document}